\documentclass{sig-alternate}
\usepackage{algorithmic}
\usepackage{algorithm}
\usepackage{subfigure}
\usepackage{graphicx}
\usepackage{epsfig}
\usepackage{cite}
\usepackage{threeparttable}
\usepackage{ntheorem}

\begin{document}
\conferenceinfo{ISPD 2011,} {March 12--15, 2011, San Jose, California, USA.} \CopyrightYear{2011} \crdata{ ACM 978-1-59593-627-1/07/0006}

\theoremstyle{plain}
\theoremheaderfont{\normalfont\bfseries}\theorembodyfont{\slshape}
\theoremsymbol{\ensuremath{\diamondsuit}} \theoremseparator{:}
\newtheorem{mytheorem}{Theorem}

\theoremsymbol{\ensuremath{\heartsuit}}
\theoremnumbering{greek}
\newtheorem{mylemma}{Lemma}

\theoremstyle{change}
\theorembodyfont{\upshape}
\theoremsymbol{\ensuremath{\ast}}
\theoremseparator{}
\newtheorem{myxample}{Example}

\theoremheaderfont{\sc}\theorembodyfont{\upshape}
\theoremstyle{nonumberplain} \theoremseparator{}
\theoremsymbol{\rule{1ex}{1ex}}
\newtheorem{myproof}{Proof}

\theoremstyle{plain} \theoremsymbol{\ensuremath{\clubsuit}}
\theoremseparator{.}
\theoremnumbering{roman}
\newtheorem{mydefinition}{Definition}

%

\pagestyle{empty} 

\title{
GLOW: A Global Router for Low-Power Thermal-Reliable Interconnect Synthesis using Photonic Wavelength Multiplexing
}

\author{
\authorblockN{
Duo Ding, Bei Yu, Ghosh Joydeep
 and
David Z. Pan\\}
\authorblockA{Dept. of ECE, The University of Texas at Austin, Austin, TX 78712\\}
\authorblockA{\authorrefmark{1}IBM T. J. Watson Research Center, Yorktown Heights, NY 10598\\}
\{ ding, dpan \}@cerc.utexas.edu}
\numberofauthors{1}
\author{
\alignauthor Duo Ding, Bei Yu and David Z. Pan\\
\affaddr{ECE Dept. Univ. of Texas at Austin, Austin, TX 78712} \\
\affaddr{\{ding, bei, dpan\}@cerc.utexas.edu}
}

{

\maketitle
\thispagestyle{empty}

\begin{abstract}
In this paper, we examine the integration potential and explore the design space of low power thermal reliable on-chip interconnect synthesis featuring nanophotonics Wavelength Division Multiplexing (WDM). With the recent advancements, it is foreseen that nanophotonics holds the promise to be employed for future on-chip data signalling due to its unique power efficiency, signal delay and huge multiplexing potential.
However, there are major challenges to address before feasible on-chip integration could be reached. In this paper, we present \emph{GLOW}, a hybrid global router to provide low power opto-electronic interconnect synthesis under the considerations of thermal reliability and various physical design constraints such as optical power, delay and signal quality. \emph{GLOW} is evaluated with testing cases derived from ISPD07-08 global routing benchmarks. Compared with a greedy approach, \emph{GLOW} demonstrates around 23\%-50\% of total optical power reduction, revealing great potential of on-chip WDM interconnect synthesis. 
\end{abstract}
\renewcommand{\thefootnote}{}
\footnote{\scriptsize This work is supported in part by Texas Advanced Research Program.}
\footnote{\scriptsize{Asia and South Pacific Design Automation Conference (ASPDAC), Jan. 30 -- Feb. 2, Sydney, Australia, 2012}}

\vspace{-0.1in}
\section{Introduction}

As semiconductor technology roadmap extends into deeper sub-micron domain, the development of future high performance low power systems faces many key challenges. Among them, VLSI interconnect plays more and more critical roles due to:
(1) growing ratio of interconnect versus gate delay;
(2) higher operating frequency and design complexity;
(3) challenging interconnect design for low power systems.

To address the interconnect challenges for future computing systems, various alternative techniques have been proposed as potential solutions (e.g.,~\cite{Cong08:ispd,Srivasatava05:iccad,Miller09:ieee}).
Among them, nanophotonics devices and interconnect attract active researches (e.g.,~\cite{Miller09:ieee, Jiang05:aip, Vlasov08:ecoc, Connor04:slip, Koo09:ieee, Young10:IEEE}) due to their unique potential to make high speed low power on-chip inteconnect.

The many recent advances in nanophotonics devices have demonstrated great on-chip integration potential in nano-scale optical modulators, photo-detectors, couplers, switches, waveguides and WDM (Wavelength Division Multiplexing) devices.
Meanwhile researches on photonics device modeling (e.g., \cite{Koo09:ieee}) and on-chip integration (e.g., \cite{Minz06:date, Shacham08:ieee, Ding09:dac, Ding09:slip, Yan09:isca, Carloni10:date}) have also introduced new opportunities and challenges to the traditional architectural and physical design methodologies.
Specifically, on-chip networks and special architecture designs have been proposed (e.g., \cite{Shacham08:ieee,Yan09:isca,Ajay09:nocs}) to enable high throughput network communication with on-chip nanophotonic WDM links.
Lately, active studies have been carried out to compensate the temperature dependence of nanophotonic devices at both fabrication level (e.g.,\cite{Lee08:oe, Guha10:oe}) and on-chip network level (e.g.,\cite{Mohamed10:islped, Li11:jetcs}) to assist the design and optimization of thermal-reliable and power-efficient optical-electrical systems-on-chips.

At physical design level, however, studies for efficient on-chip photonics interconnect synthesis have been limited. An early work by \cite{Minz06:date} employed straight line single channel optical waveguides to perform system-on-package optical routing under some timing consideration. However, physical characteristics were not considered for photonic devices. Important issues such as optical link configuration, loss, thermal reliability and signal integrity were not properly studied. In \cite{Carloni10:date}, physical-layer effects (loss, power) are applied to photonics Network-on-Chip performance evaluation. Yet without a systematic CAD environment, it is difficult to design photonics architectures with optimal performance meanwhile taking full advantage of the power budget. In \cite{Ding09:dac,Ding09:slip}, a photonic interconnect library was presented with a physical synthesis framework for low optical power routing. But thermal reliability and WDM mechanisms were not included.

\begin{figure}[b]
\label{figure1}
\vspace{-0.1in}
 \centering
\includegraphics[width=3.4in]{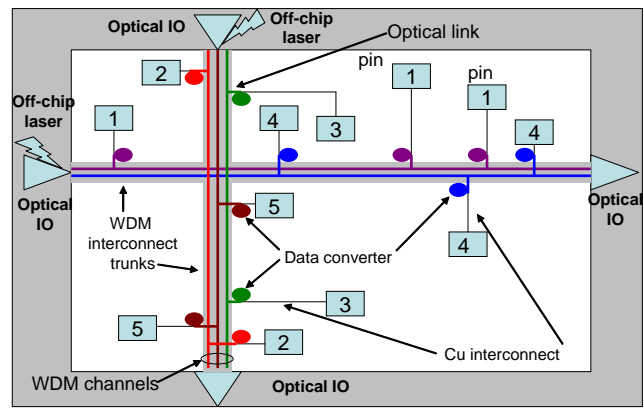}
\vspace{-0.2in}
  \caption{\scriptsize{High routing capacity interconnect using WDM}}
  \label{GLOW-basic-new}
\end{figure}

In this paper, we employ nanophotonic on-chip WDM interconnect (Fig.~\ref{GLOW-basic-new}) to achieve high density/capacity in the global routing stage. Based on device characterization and modeling, we propose \emph{GLOW}, a new hybrid global router for power-efficient thermal-reliable physical synthesis featuring WDM waveguide placement, optical channel allocations and optical-electrical data converter planning.
The rest of the paper is organized as follows: in Section~\ref{Motiv}, we motivate the WDM based optical routing problem under the critical consideration of thermal reliability and summarize the main contributions of this paper. In Section~\ref{OIL} we extend the Optical Interconnect Library (OIL) in \cite{Ding09:slip} by introducing thermal and power related models of nanophotonics devices. In Section~\ref{Section:Framework} we present an overview of our proposed CAD flow, followed by Section~\ref{CAT} and~\ref{Section:Algorithm}, in which we explain the detailed formulation and algorithms for \emph{GLOW} together with an alternative greedy approach \emph{CAT}. Section~\ref{Section:Simu} presents the results, followed by conclusion in Section~\ref{Section:Con}.

\section{Motivation and Contributions}
\label{Motiv}

With on-chip WDM providing great signal multiplexing capacity, we motivate a global router to take advantage of WDM channels under various physical design constraints such as thermal reliability and timing.
A simple scenario is illustrated in Fig.~\ref{motiv-simple}.
Given a net (A,B,C,D) to be routed with node A as the driver pin, we aim to find a global routing solution in optical-electrical domain to satisfy:

\begin{figure}[t]
\label{motiv-simple}
 \centering
\includegraphics[width=3.2in]{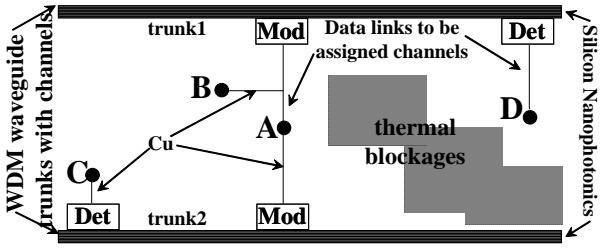}
\vspace{-0.1in}
   \caption{\scriptsize{Thermal-aware WDM routing example}}
  \label{GLOW-basic}
  \vspace{-0.15in}
\end{figure}

$\bullet$ Thermal reliability and functionality

$\bullet$ Minimal optical driving power required

$\bullet$ Signal integrity and data conversion quality


$\bullet$ Timing considerations and WDM channel utilization

$\bullet$ Legalization based on opto-electrical domain design rules

\begin{mydefinition}
    \textbf{WDM \emph{link}}: A special hybrid on-chip interconnect consisting of electrical wires and optical wavelength division multiplexing devices for large signal capacity.
\end{mydefinition}
\begin{mydefinition}
    \textbf{WDM \emph{trunk}}: The nanophotonic WDM waveguide on a WDM \emph{link} is also referred to as a \emph{trunk}.
\end{mydefinition}
\begin{mydefinition}
    \textbf{WDM \emph{channel}}: The carrier of a modulated photonic signal on a WDM \emph{trunk}. Each \emph{channel} is assigned a unique central wavelength $\lambda$. An optical signal can be transmitted on a channel if its wavelength is between $\lambda$-$\frac{1}{2}BW$ and $\lambda$+$\frac{1}{2}BW$, where $BW$ is the channel bandwidth.
\end{mydefinition}

In Fig.~\ref{motiv-simple}, thermal issue refers to the scenario for which on-chip temperature variation causes extra power loss, signal degradation or even malfunction to the nanophotnics devices, such as modulator, photo-detector and WDM waveguide. Without careful planning, an opto-electrical link could fail due to big temperature change. During global routing, regions with excessive thermal variation can be simply set as blockages to avoid. Even for regions with acceptable thermal variation, there are still trade-offs to seek between power efficiency and thermal reliability, e.g., over-optimizing optical power efficiency could result in thermal failure, whereas over-margining the thermal reliability can cause more optical power loss, especially for ring-structure resonators.

Moreover, during routing in the opto-electrical domain, there are extra timing constraints to consider such that the optical-electrical interconnects provide no worse critical path delay than the routing solutions in the electrical domain. For example in Fig.~\ref{motiv-simple}, link A$\rightarrow$B is routed with Cu interconnect while links A$\rightarrow$C, A$\rightarrow$D are partially merged with WDM trunks, meanwhile link A$\rightarrow$D takes trunk1 due to the thermal blockage between sink D and trunk2. Data links from different nets must be assigned different wavelengths (i.e., channels) when sharing the same trunk.

For high WDM channel utilization rate, sharing onto a single WDM trunk is encouraged unless timing and/or thermal conditions are violated. In this case, path A$\rightarrow$C would tend to merge with link A$\rightarrow$D onto trunk1, but it is prohibited by the long delay from trunk1 to sink C.

Last but not the least, the final routing solution needs to deliver signals strong enough to be picked up by the photo-detector, meanwhile be legalized according to design rules in both optical and electrical domains.

We summarize key contributions of this paper as follows,

$\bullet$ We propose a systematic CAD framework for on-chip WDM synthesis to co-optimize power and thermal reliability

$\bullet$ We develop a new thermal reliability models for nanophotonics devices considering WDM

$\bullet$ We formulate the optimal global routing problem with Integer Linear Programming technique

$\bullet$ We evaluate the CAD framework with various testcases derived from ISPD global routing benchmarks

\begin{table} [b]
\vspace{-0.35in}
\centering
\caption{\scriptsize{Device and interconnect model details}} \label{OILDevice:abst}
\begin{threeparttable}
\begin{scriptsize}
\begin{tabular}{|c|c|c|c|c|}

\hline
           &  footprint &  speed &      on-chip loss & E-power \\
\hline
  mod &    30X40um &     14Gb/s~\cite{Po09:oe} &   2dB & 0.7mW\\
\hline
\hline
           &  footprint &         speed &  O-det power &  E-power\\
\hline
 detector &    20X20um &      40Gb/s &     0.1mW &     1.3mW  \\
\hline
\hline
           &      delay &     optical loss &  thickness &      width \\
\hline
   WDM &    11ps/mm~\cite{Koo09:ieee} &   1.5dB/cm &      230nm &      450nm \\
\hline
\hline
           &      delay &   thickness &      width  & repeater\\
\hline
       Cu &    37ps/mm &      1um &      0.4um  & 1.4mm\\
\hline
\end{tabular}

\end{scriptsize}
\end{threeparttable}
\begin{tiny}
\begin{tablenotes}
\item
Cu interconnect on $22nm$ technology Metal5/6 with $\rho$=2.2$\mu\Omega\cdot$cm, $R_{sheet}$=0.022$\Omega$, C=2pF/cm. MOSFET models for optimal gate sizing/repeater insertion are from Metal Gate/High-K/strained-Si PTM~\cite{PTM08-Nov}.
\end{tablenotes}
\end{tiny}
\end{table}

\section{Nanophotonics Device Models}
\label{OIL}

We extend \cite{Ding09:slip} with WDM modules to analyze on-chip optical link configurations, taking into account of power, loss, timing, temperature variation and thermal reliability.

\subsection{Device Characterization}

Based on current photonics fabrication technology, optical signalling has great advantage over low-K Cu interconnect (11ps versus 37ps per mm on Metal5/6) for global nets.
Considering the delay overhead introduced by E-to-O and O-to-E data conversions, we define \emph{critical length} $L_{crit}$ as the dimension of an on-chip link above which nanophotonics yield shorter signal delay than pure Cu interconnects:

\begin{scriptsize}
\vspace{-0.1in}
\begin{equation}
T_{mod} + T_{det} + \tau_o \cdot L \leq \tau _e \cdot L
\label{ieeedt2010:critical-length}
\end{equation}
\end{scriptsize}
where $T_{mod}$ is the E-to-O modulation delay/bit and $T_{det}$ is the O-to-E photo-detection delay/bit; $\tau_o$ is signal delay per $mm$ on OWG, $\tau_e$ is the delay per $mm$ on Cu interconnect, $L$ is the length of the link. Solving Eqn. (\ref{ieeedt2010:critical-length}) gives us the range of $L$, whose lower boundary defines $L_{crit}$ value in $mm$. The devices employed in this paper are summarized in Table~\ref{OILDevice:abst}.

\subsection{Thermal Reliability for WDM}
Current on-chip WDM techniques mainly fall into the following categories: AWG (array waveguide) based, ring resonator based and thin film filter based, among which ring resonator cavity based add-drop filter techniques are most widely employed in architecture designs~\cite{Ajay09:nocs, Yan09:isca, Li11:jetcs} due to its compact footprint (potential ultra density) and demonstrated high quality factor ($Q$). However, these devices can be very sensitive to ambience temperature change.

On-chip temperature fluctuation causes the central operating frequency (wavelength) of a photonic device to drift. If such a drift results in an off-set that falls outside the range of operating bandwidth (BW), the device will degrade or even malfunction. Especially for high energy efficiency on-chip WDM devices with ring resonator structure, the quality factor $Q$~\cite{Payam02:lightwave} (defined as the energy stored in the cavity versus the energy dissipated per unit cycle) is very high and $BW$ is very narrow, thus their temperature sensitivity could lead to signal failure.
The relationships between thermal reliability, operating $BW$, quality factor $Q$ and energy efficiency are defined in Eqn.(\ref{Qfactor:equation1})-(\ref{Qfactor:equation2}).

\begin{scriptsize}
\begin{equation}
Q = \frac{\lambda_0}{\Delta \lambda_{FWHM}} = \frac{\sqrt{r_1 r_2 a} L \pi n_g}{(1 - r_1 r_2 a)\lambda_0}
\label{Qfactor:equation1}
\vspace{-0.1in}
\end{equation}
\begin{equation}
n_g(\lambda) = n_e(\lambda) - \lambda \frac{d n_e (\lambda)}{d \lambda}
\label{Qfactor:equation1.5}
\vspace{-0.1in}
\end{equation}
\begin{equation}
BW = \bigtriangleup f = \frac{f_{resonant}} {Q}
\label{Qfactor:equation2}
\end{equation}
\end{scriptsize}
where $r_1$, $r_2$, $a$, $L$ are ring geometry related parameters, $\lambda_0$ is the central working(resonant) wavelength of the ring modulator or detector. $n_e$ is a temperature dependent term, denoting the refractive index of the ring material (e.g., silicon). Therefore within a relatively small range, we can trade-off $Q$ value for thermal reliability without causing aliasing between WDM channels of trunk. Such a trade-off comes at a power loss penalty that needs to be minimized.

\begin{figure}[t]
 \centering
\includegraphics[width=2.3in]{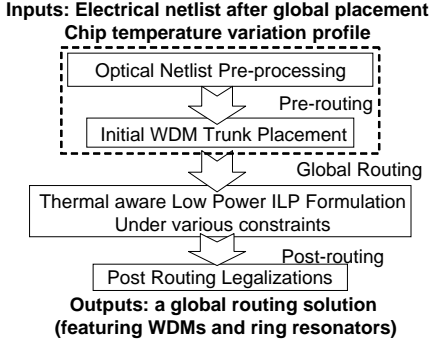}
\vspace{-0.15in}
  \caption{\small{An overview of our proposed CAD flow}}
  \label{GLOW-flow}
  \vspace{-0.15in}
\end{figure}

Based on Eqn.(\ref{Qfactor:equation1})-(\ref{Qfactor:equation2}), we establish the thermal reliability models for WDM related devices that are mainly based on cavity based components (e.g., ring resonators and ring couplers). The thermal reliability models are obtained through exhaustive temperature dependent refractive index modeling/simulation, working bandwidth characterization, power consumption/dissipation simulation and numerical methods such as Finite-difference Time-domain (FDTD) device simulations on powerful computing platforms using~\cite{Rsoft}.

\section{Overall CAD Flow}
\label{Section:Framework}

In this section, we present the overall flow of our systematic CAD framework for low power thermal-aware on-chip WDM integration, using the models from Section~\ref{OIL}.

In Fig.~\ref{GLOW-flow}, we illustrate a top level flow diagram of our proposed method, starting from a given input netlist and on-chip temperature variation profile. The flow consists of 3 major stages: a \textbf{Pre-routing} stage that prepares the optical netlist and WDM trunk placement; a \textbf{Global Routing} stage that serves as the core formulation of the WDM channel assignment problem based on various physical design constraints; and a \textbf{Post-routing} stage that further examines the legalization issues in both the optical and electrical domains. We detail the flow in Fig.~\ref{GLOW-flow} as follows.

\subsection{Netlist Pre-processing}

Netlist pre-processing step prepares the optical netlist with an initial consideration of the \emph{timing condition} which guarantees that the circuit timing does not degrade after employing nanophotonics (since each data conversion takes significant time). This step is mainly proposed to derive optical netlist test cases from existing electrical benchmarks such as ISPD07-08 global routing netlists. This step is very critical since it selects proper pins (nets or partial nets) from the electrically placed netlist to synthesize in the \textbf{Global Routing} stage. The selection is designed such that the minimal Manhattan distance of all driver-sink pairs mapped onto the optical domain is lower bounded by the \emph{critical length} $L_{crit}$. This step serves to yield \emph{non-negative timing gain} in the optical domain than in the electrical domain. This aligns well with \emph{critical length} definition and discussions in Section~\ref{OIL}. The main technique involved is described as follows,

\begin{figure}[t]
 \centering
\includegraphics[width=3.3in]{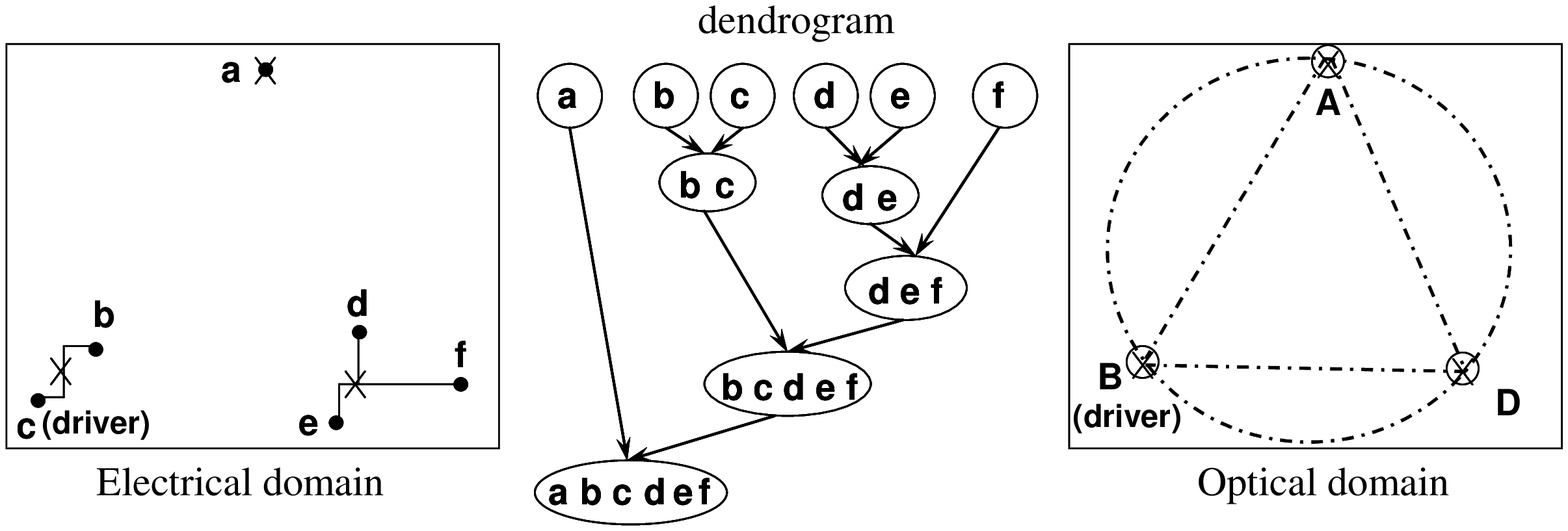}
\vspace{-0.3in}
  \caption{\scriptsize{A brief illustration of netlist pre-processing}}
  \label{GLOW-preprocess}
  \vspace{-0.15in}
\end{figure}

\textbf{Pin Clustering}: To cluster the electrically placed input netlist based on manhattan distance using hierarchical clustering method. In this case, we first construct the \emph{dendrogram} (illustrated in Fig.~\ref{GLOW-preprocess}) and then pick out the clusters satisfying the $L_{crit}$ dimension with a \emph{depth first search} on the \emph{dendrogram}. The result of this procedure is a set of clusters whose respective geometric medians are mapped to the optical domain as pseudo-pins. These pseudo-pins form the \emph{Optical Netlist}, while the rest of pins within each cluster remain on the electrical domain and are electrically interconnected to their geometric median. Therefore, only 1 O-to-E or E-to-O conversion is needed per cluster. This procedure is briefly illustrated in Fig.~\ref{GLOW-preprocess}, where $a$-$f$ are pins of certain net in the electrical netlist and $ABD$ are pseudo pins (a partial net) mapped onto the optical plane to represent clusters with edges larger than $L_{crit}$ in the \emph{dendrogram}. $B$ is the driver pin in optical domain since driver pin $c$ lies in the $bc$ cluster in electrical domain.

\subsection{Initial WDM Trunk Placement}
\label{init_placement}

Initial WDM trunk placement depend on the median of geometry distributions of optical nets in the \emph{Optical Netlist} and is carried out in a partitioned manner across the whole chip area according to Eq.~(\ref{GLOW-partition}) as a general guideline, until the total number of WDM channels is sufficient to hold the total number of optical nets/links in the netlist.
\begin{equation}
Place_{trunk^k} = med\{med[net\_i]\}^{i\; \in \; Partition^k}
\label{GLOW-partition}
\end{equation}

The partition based initial placement executes in steps:

$\bullet$ Continues for both horizontal and vertical directions

$\bullet$ Avoids over-heated regions marked as thermal blockages

$\bullet$ Partition ends when the number of WDM channels are sufficient for the total number of links in the optical netlist.

$\bullet$ Extra WDM trunks may need be added in \textbf{Post-routing}

\begin{figure}[b]
\vspace{-0.15in}
 \centering
\includegraphics[width=3.4in]{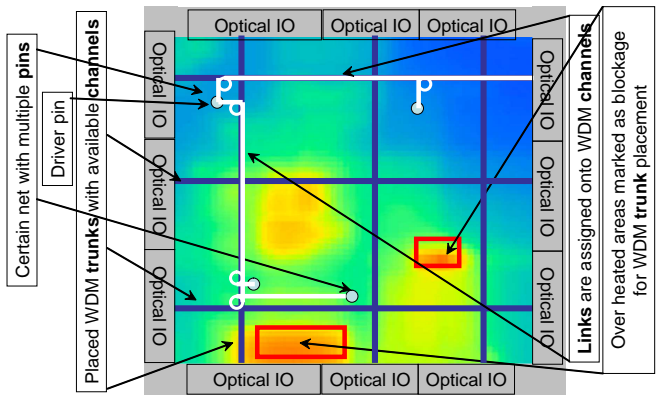}
\vspace{-0.2in}
   \caption{\scriptsize{Our WDM based global routing scenario}}
  \label{GLOW-basic}
  \vspace{-0.15in}
\end{figure}

\subsection{Thermal-aware Low Power Routing}
\label{Routing-Stage}

First, we define \emph{timing condition} as the condition that guarantees smaller signalling delay on the opto-electrical link than on Cu interconnect. This is a critical consideration since each additional O-E/E-O data conversion brings significant delay. The \emph{thermal condition}
is defined to make sure the local temperature variation does not fall out of the working range of the ring modulators. In case of a violated \emph{thermal condition}: (1) $Q$ value will be adjusted to trade-off power efficiency for thermal reliability; (2) if (1) can not be done without causing aliases between separate WDM channels, that particular region is set as a thermal blockage.
Fig.~\ref{GLOW-basic} illustrates the routing problem after \textbf{Pre-routing} stage, with off-chip laser sources whose driving power to each WDM waveguide trunk varies depending on the total number of channels assigned. To constrain the solution space for the global routing stage, we take the shortest distance route when a pin is to connect to certain WDM trunk, i.e, data converters (mod/det) are placed along WDM trunks.

For the core formulation of the \textbf{Global Routing} stage, we propose 2 algorithms: (1) global routing with low power WDM (\emph{GLOW}) as a major contribution of this paper; (2) channel assignment under thermal consideration (\emph{CAT}) as a comparison baseline. Details are in Section~\ref{CAT} and \ref{Section:Algorithm}.

\subsection{Post Routing Legalization}

This stage is mainly to resolve the cases when multiple rings are contending the same geometry location, causing design rule violations in the optical domain. For this paper, we use simple perturbation based re-routing/adjustment techniques, leaving other perspectives to detailed routing stages.

\begin{algorithm}[t]
\caption{\scriptsize{\emph{CAT}: Channel Assignment for Thermal reliability}}
 \label{Algorithm:CAT}

\begin{algorithmic}
\begin{scriptsize}
 \STATE {Input: (1) Initial WDM trunk placement}
 \STATE {Input: (2) Temperature variation profile}
 \STATE {Input: (3) Optical netlist}

 \STATE {Generate $L(link)$ as a set of all unassigned links/nets}
 \STATE {Generate $U(link)$ as a set of remaining available link resources}

 \FOR{each WDM $trunk_i$}
    \FOR {each $link_j$ in the optical netlist}
        \STATE {Calculate timing constraint on ($link_j$,$trunk_i$)}
        \STATE {Calculate thermal variation constraint on ($link_j$,$trunk_i$)}
    \ENDFOR
    \STATE {Form set $S(link_i)$ by links that satisfy \emph{Timing Condition}}
    \STATE {Sort $S(link_i)$ based on ascending \emph{Thermal Variation}}

    \STATE {Select $link_k$ $\in$ $S(link_i)$ in ascending index order}
    \WHILE {$link_k$ unassigned AND $trunk_i$ has available channels}
    \STATE {Assign $link_k$ to WDM $trunk_i$}
    \STATE {Update set $L(link)$}
    \ENDWHILE
 \ENDFOR

 \IF {set $U(link)$ = $\emptyset$ AND $L(link)$ != $\emptyset$}
 \STATE {Revise \emph{initial placement} with more WDM resources}
 \STATE {Execute \emph{CAT} on unassigned links}
 \ENDIF

\RETURN{WDM channel assignment AND optical power}
\end{scriptsize}
\end{algorithmic}
\end{algorithm}

\section{CAT Algorithm}
\label{CAT}

\emph{CAT} is designed as a greedy approach for Channel Assignment under Thermal considerations.
The basic motivation is to assign optical nets/links to WDM trunks in a sequential manner, meanwhile to combine timing and thermal-awareness constraints locally for each WDM trunk. In particular, \emph{CAT} picks all the local nets/link satisfying the timing condition and assign the least power consuming links to fill the available channels to certain WDM waveguide, then move onto the next waveguide. If eventually there are still remaining nets unassigned, then the \emph{Initial Placement} stage will be repeated to include more WDM trunks/channels.

\emph{CAT} is performed in 3 major steps: first, \emph{Initial WDM Trunk Placement}; second, \emph{Timing and Thermal Condition Calculation}; third, \emph{Greedy Channel Assignment}.

\textbf{Initial WDM Trunk Placement}: The same as Section~\ref{init_placement}, which is used for both \emph{CAT} and \emph{GLOW}.

\textbf{Timing and Thermal Condition Calculation}:
In this step, all the WDM trunks are traversed in certain order sequentially. For each trunk, timing/thermal conditions for all optical links are calculated and updated.

\textbf{Greedy Channel Assignment}:
For the channel assignment, we use a greedy heuristic method which executes in 3 phases:
\textbf{Phase1}: Form set $S(link_i)$ for WDM $trunk_i$ with the optical links that guarantee smaller signalling delay than in the electrical domain.
$S(link_i)$ is a set of link candidates to be assigned to WDM $trunk_i$.
\textbf{Phase2}: Sort the links in $S(link_i)$ with \emph{Thermal Condition} metric in ascending order.
\textbf{Phase3}: Assign links from $S(link_i)$ to $trunk_i$ in ascending order, until the total number of optical nets assigned reaches $Cmax$.
For more details of \emph{CAT}, please refer to Algorithm~\ref{Algorithm:CAT}.

\vspace{-0.1in}
\section{GLOW Routing Algorithm}
\label{Section:Algorithm}

\subsection{ILP Formulation}

To formulate the optimal optical routing problem, we represent the assignment status of channels on WDM trunks with integer binary variables (occupied or available) and use the cross-term variables to model the optical power loss introduced by WDM signal crossings, which is otherwise very difficult to accurately characterize before the channel assignment takes place. Table~\ref{GLOW:ILP} details all the variables defined.

$\bullet$ $n,m$: total number of WDM trunks in the row and column directions after initial placement, respectively.

$\bullet$ $W_{i}$: binary variables denoting the assignment status of WDM trunk $i$. If $W_i$ is 0, trunk $i$ is not unassigned any optical nets in the final routing solution, therefore will not be turned on (no input laser power from its optical IO port); if $W_i$ is 1, trunk $i$ is assigned certain nets, but may still has available channels.

$\bullet$ $W_{ij}$: binary variables numerically equal to $W_i \cdot W_j$, where $i$ $\in$ $[0,n-1]$, $j$ $\in$ $[n,n+m-1]$. 0 meaning trunk $i$ and trunk $j$ are not physically crossed; vise versa.

$\bullet$ $S_{link_k}^{trunk_i}$: binary variables, with 0 meaning link $k$ is assigned onto WDM trunk $i$.

$\bullet$ $Sum_{net_i}^{trunk_j}$: integer variables, representing the total number of optical nets assigned onto trunk $j$ in the final solution.

$\bullet$ $\lambda _{net_i}^{trunk_j}$: binary variables, 0 meaning net $i$ is assigned onto WDM trunk $j$ in the global routing; vise versa.

\begin{table} [thb]
\vspace{-0.15in}
\centering

\caption{\scriptsize{Variables/parameters in ILP formulation}}
\label{GLOW:ILP}
\begin{threeparttable}
\begin{scriptsize}
\begin{tabular}{||c|l||}
\hline
\hline
      \textbf{Name} & \textbf{Description} \\
\hline
        $P_{total}$ & total laser power consumed   \\
\hline
        $P_{loss} $ & total on-chip laser power loss    \\
\hline
        $P_{dynamic}$ & total on-chip laser power for optical signaling\\
\hline
        $P_0$ & base power consumption for a WDM trunk\\
\hline
        $P_{cross}$  &  total power loss due to trunk crossings \\
\hline
        $P_{trunk\_thm}$ & total power loss due to trunk thermal effects\\
\hline
        $P_{ring\_thm}$ & total power loss due to ring thermal effects\\
\hline
        $P_{path}$ & total power loss due to photon propagation\\
\hline
        $P_{\lambda i}$ & laser power on channel $\lambda i$ for optical signalling \\
\hline
        $P_{thm}^{ij}$ & laser power loss when trunk $i$, trunk $j$ cross\\
\hline
        $P_{trunk\_thm}^{i}$ & thermal related power loss on trunk $i$\\
\hline
        $P_{ring}^{link_i}$ & laser power loss on the rings of link $i$\\
\hline
        $W_{i}$ & BV: allocation status of trunk $i$\\
\hline
        $W_{ij}$ & BV: crossing status of trunk $i$ and trunk $j$ \\
\hline
        $S_{link_i}^{trunk_j}$ & BV: assignment status of link $i$ onto trunk $j$\\
\hline
        $Sum_{net_i}^{trunk_j}$ & IV: \# of links in net $i$ assigned to trunk $j$\\
\hline
        $\lambda_{net_i}^{trunk_j}$ & BV: assignment status of net $i$ onto trunk $j$\\
\hline
        $T_{var}^{link_i}$ & temperature variation on the rings of link $i$\\
\hline
        $C_{max}$ & channel capacity of each WDM trunk\\
\hline
        $PIN_{max}$ & max pin \# in certain net of the optical netlist\\
\hline
        $temp^{threshold}$ & temperature variation tolerance threshold\\
\hline
        $\tau_e$ & delay per unit length on Cu interconnect\\
\hline
        $\tau_o$ & delay per unit length on optical links\\
\hline
        $\tau_{conv}$ & delay overhead by data conversions\\
\hline
        $WL_e^i$ & Cu wire length on link $i$\\
\hline
        $WL_o^i$ & optical wire length on link $i$\\
\hline
        $HPWL^{link_i}$ & half parameter wire length of link $i$\\
\hline
\hline
\end{tabular}
\end{scriptsize}
\end{threeparttable}
\end{table}

We propose the following objective function for \emph{GLOW}'s thermal-aware low power routing featuring on-chip WDM:
\begin{equation}
Minimize\{P_{total}\} \;\;w.r.t\;\; W_i, \; W_{ij},\;S_{link_i}^{trunk_j}, \;\lambda_{net_i}^{trunk_j}
\label{GLOW-obj}
\end{equation}
where
\vspace{-0.1in}
\begin{equation}
P_{total} = P_{loss} + P_{dynamic}
\label{GLOW-eq1}
\end{equation}
\begin{equation}
P_{loss} = P_{cross} + P_{trunk\_thm} + P_{ring\_thm} + P_{path}
\label{GLOW-eq2}
\end{equation}
\begin{equation}
P_{cross} = \sum_{i \in [0,n-1]}^{j \in [n,n+m-1]} {W_{ij}\ast P_{thm}^{ij}}
\label{GLOW-eq3}
\end{equation}
\begin{equation}
P_{trunk\_thm} = \sum_{i}^{i \in all\;trunks}{W_i \ast P_{trunk\_thm}^{i}}
\label{GLOW-eq4}
\end{equation}
\begin{equation}
P_{ring\_thm} = \sum_{i}^{i \in all\;trunks} {\sum_{j}^{j \in all\;links} {S_{link_j}^{trunk_i} \ast P_{ring}^{link_j}}}
\label{GLOW-eq5}
\end{equation}
\begin{equation}
P_{dynamic} = \sum_{i}^{i \in all\;trunks}{\sum_{j}^{j \in all\;nets}{\lambda_{net_j}^{trunk_i}}} P_{\lambda i} + \sum_{i}{W_i P_0}
\label{GLOW-eq6}
\end{equation}

Eq.~(\ref{GLOW-obj}) above gives the objective function of \emph{GLOW} as the total power $P_{total}$ required to drive the circuit. As shown in Eq.~(\ref{GLOW-eq1}), $P_{total}$ is divided into 2 parts: the total optical power loss on chip $P_{loss}$, which is the amount of power the drivers need to compensate for the guarantee of \emph{detection conditions} on photo-detectors; and $P_{dynamic}$, the signal switching power on WDM channel carriers.

$P_{loss}$ is divided into 4 terms: waveguide crossing power, thermal related WDM trunk power, thermal related ring resonator power and the power to compensate propagation loss of on-chip waveguide.

$P_{dynamic}$ consists of 2 terms: $P_0$ is the base power consumption for each WDM trunk, it is a constant power cost when turning on a N-channel WMD trunk; the 2nd term is the switching power on all WDM channels, which is linearly proportional to the number of channels utilized. Apparently, WDM trunk multiplexing/sharing rate is to be maximized in order to avoid unnecessary $P_0$'s.

All power related terms are modeled according to Section~\ref{OIL} and Section~\ref{Section:Framework}. Table~\ref{GLOW:ILP} further details each term.

\begin{table*} [t]
\centering

\caption{Simulation result comparisons between our proposed \emph{CAT} and \emph{GLOW}}
\label{GLOW:results}
\begin{threeparttable}
\begin{scriptsize}
\begin{tabular}{|c||cccccc||cccccc|}
\hline
\textbf{Method} &     &     & \textbf{CAT} &   &      &       &       &       &   \textbf{GLOW}   &      &  &     \\
\hline
 \textbf{Optical Netlist} &  CK1 &   CK2 &      CK3 &        CK4 &    CK5 &   CK6 &  CK1 &      CK2 &       CK3 &       CK4 &     CK5  & CK6\\
\hline
\textbf{Net \#} &   35&   70 &     137 &         240 &        437 &     996 &    35&    70 &        137 &          240 &        437 &      996 \\
\hline
\textbf{Pin \#} &  95&  187 &     391 &         658 &       1357 &    2698 &    95&   187 &        391 &          658 &       1357 &     2698 \\
\hline
\textbf{Sink \#}& 60&   117 &        254 &          418 &        920 &     1702  &    60&   117 &     254 &         418 &        920 &    1702\\
\hline

\hline
\textbf{Trunk \#\tnote{a}} & 4 & 11 & 12 & 25 & 46& 138& 5& 16& 22& 40& 87& 193\\

\hline
\textbf{Channel \#\tnote{b}} & 36 & 72 & 138& 286 & 570& 1314& 35& 79 & 152& 295& 602 &1408\\

\hline
\textbf{Avg. Chan/trunk} & 9.0 & 6.55& 11.5& 11.44& 12.39& 9.52& 7.0& 4.94& 6.9& 7.38& 6.92& 7.29\\

\hline
\textbf{Total trunk-length} & 4.8 & 13.2 & 14.4 & 30 & 55.2& 165.6& 6.0& 19.2& 26.4& 48.0& 104.4& 231.6\\

\hline
\textbf{Total power\tnote{c}}& 1.45 & 4.68 &  6.81 &   13.8 &       27.26&     65.52 &  1.00 &    2.48 &     5.27 &    7.25 &       16.63 &    32.86\\

\hline
\textbf{Power reduc.\%} & -  &    -&    -  &         -  &       -  &      - &   31.0\%  & 47.0\% &  22.6\%&  47.5\% &  39.0\% &  49.8\%\\
\hline
\end{tabular}
\end{scriptsize}
\begin{tablenotes}
\begin{scriptsize}
\item [a] Each WDM trunk has up to 32 available channels at initial placement. Unassigned trunks will be turned off after routing.
\item [b] Unassigned WDM channels will be turned off (no laser input from off-chip) in the global routing stage.
\item [c] Total power consumption is normalized to the power consumed on CK1 by \emph{GLOW}.
\end{scriptsize}
\end{tablenotes}
\end{threeparttable}
\vspace{-0.2in}
\end{table*}

\vspace{0.15in}
\subsection{Physical Design Constraints}

We present the detailed mathematical formulations of various routing constraints for \emph{GLOW} as follows:

$\bullet$ Timing constraint: for each optical link, the routing solution must not result in longer signal delay than HPWL estimated delay in the electrical domain:
\begin{equation}
\begin{scriptsize}
S_{link_i}^{trunk_j} [\tau_e \ast WL_e^i  + \tau_o \ast WL_o^i  + \tau_{conv}] \leq \tau_e \ast HPWL^{link_i}
\end{scriptsize}
\end{equation}

$\bullet$ Selection constraint: to make sure each link $i$ is only assigned to one WDM trunk. For each link $i$:
\begin{equation}
\begin{scriptsize}
\sum_{j}^{j \in all\;trunks} {S_{link_i}^{trunk_j}} = 1
\end{scriptsize}
\end{equation}

$\bullet$ Channel capacity constraint: to make sure each WDM trunk does not exceed its capacity limit.
For each trunk $j$:
\begin{equation}
\begin{scriptsize}
\sum_{i}^{i \in all\;nets} {\lambda _{net_i} ^{trunk_j}} \leq Cmax
\end{scriptsize}
\end{equation}

$\bullet$ Detection constraint: the final optical power at each sink on each link must be large enough to be detected.

$\bullet$ Thermal constraint: for each link (pair of pins from source to sink), local temperature variation be upper bounded by $temp^{threshold}$ to avoid performance degradation or malfunction.
For each link $i$ and trunk $j$:
\begin{equation}
\begin{scriptsize}
S_{link_i}^{trunk_j} \ast T_{var}^{link_i} \leq temp^{threshold}
\end{scriptsize}
\end{equation}

\begin{algorithm}[t]
\caption{\scriptsize{\emph{GLOW}: Global Routing for Low Power WDM}}
 \label{Algorithm:GLOW}

\begin{algorithmic}
\begin{scriptsize}
 \STATE {Input: (1) Initial WDM trunk placement}
 \STATE {Input: (2) Temperature variation profile}
 \STATE {Input: (3) Optical netlist}

 \FOR {each WDM $trunk_i$}
    \FOR {each optical $link_j$}
        \STATE {Calculate $P_{cross}$, $P_{WDM\_thermal}$, $P_{ring\_thermal}$}
        \STATE {Calculate $P_{dynamic}$}
        \STATE {Update \emph{Timing Constraint}, \emph{Thermal Constraint}}
    \ENDFOR
 \ENDFOR

 \STATE {Invoke ILP solver}
\RETURN{WDM channel assignment AND optical/laser power}
\end{scriptsize}
\end{algorithmic}
\end{algorithm}

$\bullet$ Binary/Integer variable constraints: since $W_{ij}$ and $\lambda_{net_i}^{trunk_j}$ are introduced to eliminate non-linear terms, the following constraints must be enforced:
\begin{scriptsize}
\begin{equation}
\label{var_cross}
2 W_{ij} \leq W_{i} + W_{j} \leq 1 + W_{ij}
\end{equation}
where $i$ $\in$ $[0,n-1]$, $j$ $\in$ $[n,n+m-1]$
\begin{equation}
\label{var_lambda}
\frac {(2 \sum _{k}^{k \in net_i} {S_{link_k}^{trunk_j}} - 1)}{2 PINmax} \leq \lambda_{net_i}^{trunk_j} \leq 2 \sum _{k}^{k \in net_i} {S_{link_k}^{trunk_j}}
\vspace{-0.0in}
\end{equation}
\begin{equation}
\label{var_W}
\frac{(2 \sum_{i = 1}^{all\;nets} {\lambda_{net_i}^{trunk_j}} - 1)}{2 Cmax} \leq  W_j \leq 2 \sum_{i = 1}^{all \;\; nets} {\lambda_{net_i}^{trunk_j}}
\vspace{-0.0in}
\end{equation}
\end{scriptsize}

Here Equation(\ref{var_lambda}) and (\ref{var_W}) are enforced for two-fold reasons: (1) we are able to calculate the number of optical nets assigned to certain WDM trunk via optical link related variables; (2) to introduce non-linear relation between $\lambda_{net_i}^{trunk_j}$ and $S_{link_k}^{trunk_j}$ under ILP formulation. For this part an intermediate term $Sum_{net_i}^{trunk_j}$ is introduced by Equation(\ref{var_sum}):
\begin{equation}
\begin{scriptsize}
\label{var_sum}
Sum_{net_i}^{trunk_j}  = \sum_{k}^{k \in net_i}{S_{link_k}^{trunk_j}}
\end{scriptsize}
\end{equation}

Equation(\ref{var_lambda})(\ref{var_W})(\ref{var_sum}) together make sure that if $Sum_{net_i}^{trunk_j}$ = 0, then $\lambda _{net_i}^{trunk_j}$ = 0; if $Sum_{net_i}^{trunk_j}$ $>$ 0, then $\lambda_{net_i}^{trunk_j}$ = 1. They serve as a binary comparator meanwhile still preserve the linear programming formulation.
Algorithm~\ref{Algorithm:GLOW} summarizes the pseudo-codes of the main steps of \emph{GLOW}.

\section{Simulation and Testing}
\label{Section:Simu}

\textbf{Benchmarks and Simulation Setups}:
In Table~\ref{GLOW:results} we list 6 optical benchmarks: CK1-6, with net number ranging from 35 to 996. These test cases are derived from IPSD07-08 global routing contest benchmarks (with over 100K nets) by: (1) up-scaling the chip dimension into centimeter scale; (2) employing our proposed \emph{Optical Netlist Pre-processing} techniques to generate optical netlists. Considering the limited integration volume of current on-chip WDM nanophotonics, the sizes of these testing netlists are suitable.

For the hierarchical clustering procedure, $L_{crit}$ is set to 3.7mm for centimeter-scale chips. We assume all the inserted ring resonators are legalized and initially thermally tuned. The on-chip thermal variation profiles are randomly generated based on measured data of real processor chips. The tolerance threshold $temp^{th}$ of the maximal range of temperature variation is set to between 15 to 20 degrees, as hard constraints in our problem formulation. Corresponding wavelength off-set sensitivity of the WDM interconnect is set to 0.12nm/degree C. For the WDM trunk initial placement, we use 32-channel WDM trunks to start with, then run the proposed global routing algorithms on 3.0GHz Linux workstations with 8GB memories.

\textbf{Result and Analysis}:
In Table~\ref{GLOW:results}, we show simulation results of \emph{CAT} and \emph{GLOW}, with total power consumption normalized to the power value that \emph{GLOW} gives on CK1.
Compared with \emph{CAT}, \emph{GLOW} demonstrates around 23\%-50\% of total power reductions on CK1-6, respectively.

Reasons of such improvement are mainly two-fold: first, \emph{CAT} only searches for local optimal solutions and assign optical nets/links to WDM trunks in a sequential/local manner, while \emph{GLOW} aims at global optimal solution with mathematical programming techniques; second, \emph{CAT} is not aware of the waveguide crossing power, nor does it consider the thermal related ring resonator power-reliability trade-off globally; on the other hand, the ILP formulation of \emph{GLOW} allows us to model all the key factors of optical power.

Also in Table~\ref{GLOW:results} we show the WDM channel/trunk allocation of \emph{CAT} and \emph{GLOW} on CK1-6. We see that compared with \emph{GLOW}, \emph{CAT} assigns fewer number of WDM trunks, resulting in a slightly higher number of average WDM channels per trunk and shorter total length of on-chip WDM waveguide. \emph{GLOW}, however, works by assigning WDM trunks/channels across the chip aiming at the global solution of power consumption minimization under given thermal reliability requirements. This helps \emph{GLOW} to bring down the total power at the cost of some extra OWG wirelength. This is acceptable since the fabrication costs of straight OWGs are relatively low meanwhile there are resources on the silicon layer for the nanophotonics integration.

In some few cases when there are no feasible solutions exist, the ILP formulation will not return valid WDM channel/trunk allocation strategy and the WDM trunk initial placement must be adjusted (by adding more trunks). In this paper, such adjustments are carried out in a progressive and heuristic manner until feasible integer solutions are found.
With accelerated ILP, \emph{GLOW} manages to locate the optimal solutions for the 6 optical netlists in about 0.2 to 0.9 hours. Such run-time is acceptable as the optical routing problem size is fairly small, i.e., only the top global nets/pins are mapped into the optical domain while the rest nets are routed in the electrical domain.

\section{Conclusion}
\label{Section:Con}

In this paper we explored the synthesis of on-chip nanophotonic WDM interconnects and presented \emph{GLOW}, a low power optical router under thermal-reliability considerations. It is evaluated on various testing cases showing significant optical power reduction compared with a baseline greedy method. We believe a lot of future research can be done to co-optimize the CAD and the nanophotonics technologies in the physical design area. 
}

\tiny
\bibliographystyle{unsrt}
\bibliography{GLOW-2010-short}

\begin{thebibliography}{10}

\bibitem{Cong08:ispd}
M.-C.~Frank Chang et~al.
\newblock {RF Interconnects for Communications On-Chip}.
\newblock In {\em Proc. Int. Symp. on Physical Design}, 2008.

\bibitem{Srivasatava05:iccad}
Navin Srivastava et~al.
\newblock {Performance Analysis of Carbon Nanotube Interconnects for VLSI
  Applications}.
\newblock In {\em ICCAD}, 2005.

\bibitem{Miller09:ieee}
David A.~B. Miller.
\newblock {Device Requirement for Optical Interconnects to Silicon Chips}.
\newblock In {\em Proc. of the IEEE, Special Issue on Silicon Photonics}, 2009.

\bibitem{Jiang05:aip}
Yongqiang Jiang et~al.
\newblock {80-micron Interaction Length Silicon Photonic Crystal Waveguide
  Modulator}.
\newblock In {\em Applied Physics Letters}, 2005.

\bibitem{Vlasov08:ecoc}
Yurii Vlasov.
\newblock {Silicon Photonics for Next Generation Computing Systems}.
\newblock In {\em European Conference on Optical Communications}, 2008.

\bibitem{Connor04:slip}
Ian. O'Connor.
\newblock {Optical Solutions for System-Level Interconnect}.
\newblock In {\em Proc. System Level Interconnect Prediction}, 2004.

\bibitem{Koo09:ieee}
Kyung-Hoae Koo et~al.
\newblock {Compact Performance Models and Comparisons for Gigascale On-Chip
  Global Interconnect Technologies}.
\newblock In {\em IEEE Trans. on Electron Devices}, 2009.

\bibitem{Young10:IEEE}
Ian~A. Young et~al.
\newblock {Optical I/O Technology for Tera-Scale Computing}.
\newblock In {\em IEEE J. Solid-State Circuits}, 2010.

\bibitem{Minz06:date}
Jacob~R. Minz et~al.
\newblock {Optical Routing for 3D System-on-Package}.
\newblock In {\em Proc. Design, Automation and Test in Eurpoe}, 2006.

\bibitem{Shacham08:ieee}
Assaf Shacham et~al.
\newblock {Photonic Networks-on-Chip for Future Gene- -ration Chip
  Multiprocessors}.
\newblock In {\em IEEE Trans. on Computers}, 2008.

\bibitem{Ding09:dac}
Duo Ding et~al.
\newblock {O-Router: An Optical Routing Framework for Low Power On-Chip Silicon
  Nano-photonics Integration}.
\newblock In {\em Proc. Design Automation Conf.}, 2009.

\bibitem{Ding09:slip}
Duo Ding et~al.
\newblock {OIL: A Nanophotonic Optical Interconnect Library for a New Photonic
  Networks-on-Chip Architecture}.
\newblock In {\em Proc. System Level Interconnect Prediction}, 2009.

\bibitem{Yan09:isca}
Yan Pan et~al.
\newblock {Firefly: Illuminating Future Network-on-Chip with Nanophotonics}.
\newblock In {\em Proc. Int. Symp. on Computer Architecture}, 2009.

\bibitem{Carloni10:date}
Johnnie Chan et~al.
\newblock {PhoenixSim: A Simulator for Physical-Layer Analysis of Chip-Scale
  Photonic Interconnection Networks}.
\newblock In {\em Proc. Design, Automation and Test in Eurpoe}, 2010.

\bibitem{Ajay09:nocs}
Ajay Joshi et~al.
\newblock {Silicon-Photonic Clos Networks for Global On-Chip Communication}.
\newblock In {\em Int. Synp. on Networks-on-Chip}, 2009.

\bibitem{Lee08:oe}
Jong-Moo Lee et~al.
\newblock {Controlling Temperature Dependence of Silicon Waveguide Using Slot
  Structure}.
\newblock In {\em Optics Express}, 2008.

\bibitem{Guha10:oe}
Biswajeet Guha et~al.
\newblock {CMOS-Compatible Athermal Silicon Microring Resonators}.
\newblock In {\em Optics Express}, 2010.

\bibitem{Mohamed10:islped}
Moustafa Mohamed et~al.
\newblock {Power-Efficient Variation-Aware Photonic On-Chip Network
  Management}.
\newblock In {\em Int. Symp. on Low Power Electronics Design}, 2010.

\bibitem{Li11:jetcs}
Zheng Li et~al.
\newblock {IRIS: A Hybrid Nanophotonic Network Design for High-Performance and
  Low-Power On-Chip Communication}.
\newblock In {\em J. on Emerging Technologies in Computing Systems}, 2011.

\bibitem{Po09:oe}
Po~Dong et~al.
\newblock {Low Vpp, Ultralow-energy, Compact, High-speed Silicon Electro-Optic
  Modulator}.
\newblock In {\em OPTICS EXPRESS}, 2009.

\bibitem{PTM08-Nov}
{Predictive Technology Model, http://ptm.asu.edu}.

\bibitem{Payam02:lightwave}
Payam Rabiei et~al.
\newblock {Polymer Micro-Ring Filters and Modulators}.
\newblock In {\em J. of Lightwave Technology}, 2002.

\bibitem{Rsoft}
{Rsoft is a CAD Suite for Photonics Device Simulation}.

\end{thebibliography}

\end{document}